\documentclass[runningheads]{llncs}
\usepackage[T1]{fontenc}
\usepackage{graphicx}
\usepackage{hyperref}
\usepackage{url}
\usepackage{times}

\usepackage{soul}
\usepackage{url}

\usepackage{bm}
\usepackage{multirow}
\usepackage{multicol}
\usepackage{stfloats}
\usepackage{caption}
\usepackage{subfigure}
\usepackage{enumitem}
\usepackage{booktabs}
\usepackage{amsmath}
\usepackage{amsfonts}
\usepackage{wrapfig}
\usepackage{mathtools}

\DeclarePairedDelimiter\ket{\lvert}{\rangle}
\DeclarePairedDelimiterX\braket[2]{\langle}{\rangle}{#1 \delimsize\vert #2}
\begin{document}
\title{QuanGCN: Noise-Adaptive Training for Robust Quantum Graph Convolutional Networks}
%
%
\author{Kaixiong Zhou\inst{1} \and
Zhenyu Zhang\inst{2} \and Shengyuan Chen\inst{3} \and Tianlong Chen\inst{2} \and Xiao Huang\inst{3} \and Zhangyang Wang\inst{2} \and Xia Hu\inst{1}}
\authorrunning{Kaixiong Zhou et al.}
%
\institute{Rice University, Houston, Texas 77054, \{kaixiong.zhou, xia.hu\}@rice.edu \and
The University of Texas at Austin, Austin, Texas 78712, \{zhenyu.zhang, tianlong.chen, atlaswang\}@utexas.edu \and
The Hong Kong Polytechnic University, Hong Kong, \{shengyuan.chen@connect.polyu.hk, xiaohuang@comp.polyu.edu.hk\}}
\maketitle              
%
\begin{abstract} 
Quantum neural networks (QNNs), an interdisciplinary field of quantum computing and machine learning, have attracted tremendous research interests due to the specific quantum advantages. Despite lots of efforts developed in computer vision domain, one has not fully explored QNNs for the real-world graph property classification and evaluated them in the quantum device. To bridge the gap, we propose quantum graph convolutional networks (QuanGCN), which learns the local message passing among nodes with the sequence of crossing-gate quantum operations. To mitigate the inherent noises from modern quantum devices, we apply sparse constraint to sparsify the nodes' connections and relieve the error rate of quantum gates, and use skip connection to augment the quantum outputs with original node features to improve robustness. The experimental results show that our QuanGCN is functionally comparable or even superior than the classical algorithms on several benchmark graph datasets. The comprehensive evaluations in both simulator and real quantum machines demonstrate the applicability of QuanGCN to the future graph analysis problem\footnote{The Conference Quantum Techniques in Machine Learning (QTML), 2022.}.

\end{abstract}

\section{Introduction}
Quantum computing is emerging as a powerful computational paradigm~\cite{cao2019quantum,kandala2017hardware,biamonte2017quantum,farhi2014quantum,harrow2009quantum,rebentrost2014quantum}, showing impressive efficiency in tackling traditionally intractable problems,   
including cryptography~\cite{shor1999polynomial} and database search~\cite{grover1996fast}.
With trainable weights in quantum circuits, quantum neural networks (QNNs), such as quantum convolution~\cite{henderson2020quanvolutional} and quantum Boltzmann machine~\cite{amin2018quantum}, have achieved speed-up over classical algorithms in machine learning tasks, including metric learning~\cite{lloyd2020quantum} and principal component analysis~\cite{lloyd2014quantum}.


Despite the successful outcomes in processing structured data (e.g., images~\cite{wang2021roqnn,wang2021roqnn}), QNNs are rarely explored for the graph analysis. Graphs are ubiquitous in the real-world systems, such as biochemical molecules~\cite{dai2016discriminative,duvenaud2015convolutional,zhou2021multi} and social networks~\cite{hamilton2017inductive,velivckovic2017graph,zhou2021dirichlet,chen2022bag}, where graph convolutional networks (GCN) has become the de-facto-standard analysis tool~\cite{kipf2017semi}. By passing the large-volume and high-dimensional messages along edges of the underlying graph, GCN learns the effective node representations to predict graph property. Given the time-costly graph computation, QNNs could provide the potential acceleration for via the superposition and entanglement of quantum circuits. 

\begin{figure}
    \centering
    \includegraphics[width=\textwidth]{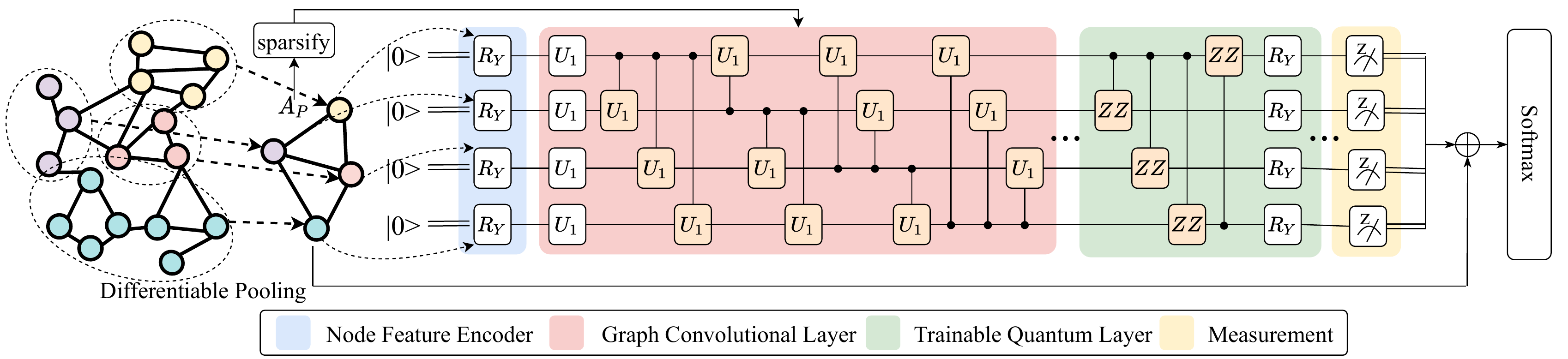}
    \vspace{-10pt}
    \caption{The classical-quantum hybrid framework of QuanGCN for graph classification problem, which encoder pools the input molecule and encodes the clustered node features with qubits, graph convolutional layer inserts quantum gate between every pair of qubits to approximate classical message passing, and measurement circuits read out graph representation to estimate labels.}
    \label{fig:system}
\end{figure}

However, the existing quantum GCN algorithms cannot be directly applied for the real-world graph analysis. They are either developed for the image recognition or quantum physics~\cite{zheng2021quantum,verdon2019quantum}, or are only the counterpart simulations in the classical machines~\cite{dernbach2018quantum,beer2021quantum}. Even worse, most of them do not provide the open-source implementations. 
To tackle these challenges, as shown in Figure~\ref{fig:system}, we propose quantum graph convolutional networks (QuanGCN) towards the graph classification tasks in real-world applications. Specifically, we leverage differentiable pooling layer to cluster the input graph, where each node is encoded by a quantum bit (qubit). The crossing-qubit gate operations are used to define the local message passing between nodes. QuanGCN delivers the promising classification accuracy in the real quantum system of IBMQ-Quito.

The existing quantum devices suffer from  non-negligible error rate in the quantum gates, which may lead to the poor generation of QuanGCN. To mitigate the noisy impact, we propose to apply the  sparse constraint and skip connection. While the sparse constraint sparsifies the pooled graph and reduces the scales of crossing-qubit gate operations, the skip connection augments the quantum outputs with the classical node features. In summary, we make the following three contributions: (1) The first QuanGCN to address the real-world graph property classification tasks; (2) Two noise mitigation techniques used to improve model's robustness; (3) The extensive experiments in validating the effectiveness of QuanGCN compared with classical algorithms. 

\section{Methodology}
We represent an undirected graph as $G = (A, X)$, where $A\in\mathbb{R}^{n\times n}$ denotes adjacency matrix, and $X\in\mathbb{R}^{n\times d}$ denotes feature matrix, $n$ is the number of nodes, and the $i$-th row $x_i$ at matrix $X$ is the feature vector of node $v_i$. The goal of graph classification task is to predict label of each graph (e.g., biochemical molecule property). Specifically, given a set of graphs $\{(G_1, y_1), (G_2, y_2), \cdots\}$ where $y_g$ is the corresponding label of graph $G_g$, we learn the representation vector $h_g$ to classify the entire graph: $y_g = f(h_g)$.

\subsection{Preliminary of Graph Convolutional Networks}
\label{sec: GCN}
The node embedding $x^{(l)}_i\in\mathbb{R}^{d}$ at the $l$-th layer of a graph neural networks is generally learned according to~\cite{hamilton2017inductive,zhou2019auto,sun2022gppt}:
\begin{equation}
    \label{equ:GCN}
    x^{(l)}_i = \mathrm{Aggregate}(\{a_{ij}x^{(l-1)}_jW^{(l)}: j\in\mathcal{N}(i)\cup v_i \}).
\end{equation}
$\mathcal{N}(i)$ denotes the set of neighbors adjacent to node $v_i$; $a_{ij}$ denotes the edge weight connecting nodes $v_i$ and $v_j$, which is given by the $(i,j)$-th element at matrix $A$; $\mathrm{Aggregate}$ denotes the permutation-invariant function to aggregate the neighborhood embeddings, and combine them with the node itself, i.e., $v_i$. The widely-used aggregation modules include sum, add, and mean functions. $W^{(l)}\in\mathbb{R}^{d\times d}$ is trainable projection matrix. 
Considering a $L$-layer GCN, READOUT function (e.g., sum or mean) collects all the node embeddings from the final iteration to obtain the graph representation: $h_g = \mathrm{READOUT}(x^{(L)}_i | i=1, \cdots, n)$, which is used for the graph classification task.
\subsection{Quantum Graph Convolutional Networks}
We propose QuanGCN to realize the graph representation learning in the classical-quantum hybrid machine in Figure~\ref{fig:system}, which is consisted of three key components.

\paragraph{\textbf{Pooling Quantum State Encoder.}} This state is responsible to encode the classical node features into quantum device, where each node is is represented by a qubit. Since the existing quantum machines have a limited number $q$ of qubits, it is intractable to encode the graphs with thousands of nodes. In this work, we leverage a differentiable pooling module to cluster each graph to a fixed $q$-node coarsened graph. Specifically, let $S=
\mathrm{Pool}(A, X) \in \mathbb{R}^{n\times q}$ denote the clustering matrix, where $\mathrm{Pool}$ is trainable module of MLP or other advanced pooling networks~\cite{gao2019graph,ying2018hierarchical,zhou2020towards}. Each row at matrix $S$ indicates the probabilities of a node being pooled to the $q$ clusters. We could obtain the adjacency matrix and node features of coarsened graph as follow:
\begin{equation}
\label{eq:pool}
A_p = S^T A S \in \mathbb{R}^{q\times q}; \quad X_p = S^T X \in \mathbb{R}^{q\times d}.
\end{equation}
We encode the node features of pooled graph with rotation gates. To simplify the analysis and without loss of generality, we assume node feature dimension to be $d=1$. The high dimensional feature could be encoded by repeating the process or using complicated quantum gates. To be specific, let $\ket{\phi}=\ket{0,...,0}$ denote the ground statevector in $q$-qubit quantum system. The computation
on a quantum system is implemented by a sequence of parameterized quantum gates on statevector $\ket{\phi}$. Parameterized by node features, we use a sequence of  $\mathsf{R_y}$ gates to encode the pooled graph as: $\ket{\phi} = \mathsf{R_y}(x_p, [q])\cdots \mathsf{R_y}(x_1, [1])\cdot\ket{\phi}$. $\mathsf{R_y}(x_i, [i])$ denotes the single-qubit quantum gate rotating the $i$-th qubit along $y$-axis, at which the rotation angle is characterized by node feature $x_i$ (i.e., the $i$-th row in $X_p$). In other word, the node features are memorized in the quantum system by the rotation angles of quantum states. 

\paragraph{\textbf{Quantum Graph Convolution.}} As defined in Eq.~\eqref{equ:GCN}, the representation of node $v_i$ is computed by incorporating the self-loop information and aggregating the neighborhood embeddings. In the quantum counterpart, we use the quantum gates of $\mathsf{U1}$ and $\mathsf{CU1}$ to model the self-loop and node-pairwise message passing, respectively:
\begin{equation}
\label{eq: quangcn}
    \ket{\phi} = \bigcap_{j=1 \& j\neq i}^{q}\mathsf{CU1}(\hat{a}_{ij}, [j, i]) \cdot \mathsf{U1}(\hat{a}_{ii}, [i])\cdot \ket{\phi}.
\end{equation}
$\mathsf{CU1}(\hat{a}_{i,j}, [j, i])$ is a two-qubit quantum gate, where the ordered pair $[j, i]$ means the $j$-th quantum circuit is a control qubit and the $i$-th one is target qubit. The unitary operation working on qubit $i$ is parameterized edge weight $\hat{a}_{ij}$, i.e., the $(i,j)$-th element at matrix $A_p$. Symbol $\bigcap$ denotes the sequential gate operations. $\mathsf{U1}(\hat{a}_{ii}, [i])$ is a single-qubit quantum gate working on qubit $i$, which is parameterized by the self-loop weight $\hat{a}_{ii}$. By applying Eq.~\eqref{eq: quangcn} to all the qubits, we model the quantum message passing between node pairs. A following trainable quantum layer is then used as shown in Figure~\ref{fig:system}.

\paragraph{\textbf{Measurement.}} After $L$ layers of quantum graph convolutions, we measure the expectation values with $\mathsf{Pauli}$-$\mathsf{Z}$ gate and obtain the classical float value from each qubit. The measurements are concatenated to predict the graph labels as described in Section~\ref{sec: GCN}.

\subsection{Noise Mitigation Techniques}
In the real quantum systems, noises
often appear due to the undesired gate operation. To mitigate noise in our QuanGCN, we propose to apply the following two techniques. 

\paragraph{\textbf{Pooling sparse constraint.}} The operation error generally increases with the number of included quantum gates. One of the intuitive solutions to relieve noise is to sparsify the adjacency matrix of pooled graph, where most of the edge weights are enforced to be zero. In this way, the applied quantum gate $\mathsf{CU1}$ or $\mathsf{U1}$ could be treated as an identity operation, which rotates the target qubit with the angle of zero. Specifically, we adopt the entropy constraint to learn the sparse adjacency matrix: $\mathcal{L}_{\mathrm{sparse}}= -\sum_{i}\sum_{j}\hat{a}_{ij}\log \hat{a}_{ij}$, which is co-optimized with the graph classification loss.

\paragraph{\textbf{Skip connection.}} We mitigate the quantum noise from the architectural perspective by introducing the skip connection. In Figure~\ref{fig:system}, we concatenate the quantum measurements with the input classical features, which is not sensitive to the quantum noise.

\section{Experimental Setting and Results}
\paragraph{\textbf{Dataset.}} We adopt four graph datasets, including two bioinformatic datasets of ENZYMES, PROREINS~\cite{borgwardt2005protein,feragen2013scalable}, and two social network datasets of MUTAG and IMDB-BINARY~\cite{dobson2003distinguishing}. They contain 600, 1113, 188, and 1000 graphs, respectively. 

\paragraph{\textbf{Implementations.}} We adopt the classical baselines of MLP, simplified graph convolutions (SGC)~\cite{wu2019simplifying}, GCN, and the graph pooling method of Diffpool~\cite{ying2018hierarchical}. SGC uses MLP to learn the node presentations based on the preprocessed node features. 
For QNN algorithms, besides QuanGCN, we include quantum MLP (QuanMLP) and quantum SGC (QuanSGC), where their MLP layers are replaced by the quantum layer of  $\mathsf{U3CU3}$. The numbers of qubits and graph convolutional layers are set to $4$ and $2$, respectively.

\paragraph{\textbf{Comparison of classical and quantum neural networks.}} We compare the classification accuracies in Table~\ref{tab:classical_quantum}, where the mean results and standard variances are reported with $10$ random runs. It is observed that our QuanGCN obtains the comparable or even superior results than the cassical algorithms, while generally outperforming QuanMLP and QuanSGC on the benchmark datasets. These results validate the effectiveness of quantum graph convolution in dealing with the graph data. By modeling the time-expensive message passsing in the efficient quantum device, QuanGCN provides the potential speed-up over the classical algorithms.  Similar to other QNNs, QuanGCN is accompanied with higher variance due to the indeterminate quantum operations.

\begin{table}[h]
    \centering
    \begin{tabular}{c|c|cccc}
     \toprule
       Frameworks  & Methods & ENZYMES & MUTAG & IMDB-BINARY  & PROTEINS  \\
       \hline
       \multirow{4}*{Classical} & MLP &32.17±1.77 &78.95±0.00 & 70.10±1.10 & 70.37±0.76
 \\
       & SGC &49.00±4.66 &84.21±0.00 &69.70±3.13 & 72.66±1.72 \\
       & GCN &52.33±3.44 &82.63±2.54 & 70.40±1.90 &  71.65±1.26
 \\
        & DiffPool &50.00±3.60 &  78.95±6.08 & 71.90±1.91  & 69.63±1.64 \\
         \hline
         \multirow{3}*{Quantum} & QuanMLP-w/o noise &31.67±1.57 &78.95±0.00 & 72.10±0.99
 & 67.68±2.34\\
          
         & QuanSGC-w/o noise &37.83±4.52 &80.53±4.33 & 69.90±2.92  & 67.86±0.94 \\
         
         & QuanGCN-w/o noise&50.00±6.57  &83.16±5.44 & 71.10±2.77  & 70.00±2.77 \\
       
    \bottomrule
    \end{tabular}
    \caption{Graph classification accuracies in percent; QNNs are trained and inferred in GPUs without inserting quantum noise.}
    \label{tab:classical_quantum}
\end{table}

          
         
       

\vspace{-30pt}
\paragraph{\textbf{Testing in quantum simulator and real machine.}} In Table~\ref{tab:inference}, we deploy the above well-trained QNNs in Qiskit simulator and quantum computer of IBMQ-Quito to evaluate their inference performances. Since the inference in real quantum computer has to pay plenty of queuing time, we test QNNs only once in the real device. Comparing with the inference performances in GPUs (i.e., in Table~\ref{tab:classical_quantum}), QNNs generally have lower accuracies due to the high error rates existing inherently in the quantum devices. Notably, QuanGCN instead obtains the better performances. One of the possible reasons is due to the graph pooling, which highly reduces the crossing-qubit gate usages and the resultant noises. The quantum graph convolution over the pooled graph provides the more informative encoding for the underlying graph structure.
\begin{table}[h]
    \centering
    \begin{tabular}{c|c|ccccc}
     \toprule
       Frameworks  & Methods & ENZYMES & MUTAG & IMDB-BINARY  & PROTEINS  \\
       \hline
         \multirow{3}*{Simulator} 
          & QuanMLP-noise &20.50±5.21 &62.11±9.54 & 70.90±5.02  & 48.07±9.95\\
         
         & QuanSGC-noise &22.83±4.91 &61.05±13.18 & 73.50±4.93 &  50.55±10.86 \\
        
         & QuanGCN-noise &78.67±5.76 &88.95±5.23 & 76.30±3.74 & 74.77±2.49\\
         \hline
         \multirow{3}*{Real QC} 
          & QuanMLP-noise &18.33 &63.16 &  54.00   & 40.37\\
         
         & QuanSGC-noise &21.67 &63.16 & 65.00 & 59.63\\
        
         & QuanGCN-noise &83.33 &84.21 & 78.00 & 78.90 \\
       
    \bottomrule
    \end{tabular}
    \caption{Inference results of graph classification accuracies in the environments of Qiskit simulator and real quantum computer.}
    \label{tab:inference}
\end{table}

\paragraph{\textbf{Noise mitigation results.}} To address the inherent noisy impacts, we apply the skip connection to all the QNNs, and use the sparse constraint to regularize the graph pooling in QuanGCN. We compare them with one popular noise cancellation baseline~\cite{wang2021roqnn}, which randomly inserts quantum gates during model training to improve robustness. The comparison results in Tabel~\ref{tab:denoise} show that the technique of skip connection is consistently effective to mitigate noise in all models. In QuanGCN, the combination of skip connection and sparse constraint obtains the best noise mitigation performances.

\begin{table}[h]
    \centering
    \begin{tabular}{c|c|ccccc}
     \toprule
       Frameworks  & Methods & ENZYMES & MUTAG & IMDB-BINARY & PROTEINS  \\
       \hline
         \multirow{2}*{QuanMLP-noise} 
          & Random injection &22.33±6.15 &60.53±14.09 & 56.70±5.21  & 50.37±6.26\\
         
         & Skip connection &27.83±2.09 &63.68±11.22 & 72.20±1.03 & 64.22±6.27\\
        
         \hline
         \multirow{2}*{QuanSGC-noise} 
          & Random injection &20.50±6.19 &64.74±6.59 & 61.80±8.46 & 51.19±6.78\\
         
         & Skip connection &29.00±5.45 &67.37±14.42 & 71.30±1.77 &  68.07±4.98 \\
         \hline
         \multirow{4}*{QuanGCN-noise} 
         & Random injection &35.17±18.53 &59.47±30.09 & 57.50±14.18 & 63.12±8.02 \\
         & Skip connection &49.33±9.27 &86.84±3.72 & \bf{71.90±2.42} & 72.02±2.42 \\
         & Sparse &41.67±16.52 &63.16±20.61 & 64.30±9.9  & 60.46±7.90\\
         & Skip + Sparse &\bf{49.83±8.22} &\bf{86.84±6.68} & 70.40±2.01  & \bf{73.30±1.91}\\
       
    \bottomrule
    \end{tabular}
    \caption{Quantum noise mitigation results with skip connection and sparse constraint.}
    \label{tab:denoise}
\end{table}

\section{Conclusion}
In this work, we propose and implement QuanGCN towards addressing the graph property classification tasks in the real-world applications. 
To mitigate the noisy impact in the real quantum machine, we propose techniques of skip connection and sparse constraint to improve model's robustness. The extensive experiments on the benchmark graph datasets demonstrate the potential advantage and applicability of quantum neural networks to the graph analysis, which is a new probem introduced to quantum domain.

\clearpage

\bibliographystyle{unsrt}
\bibliography{gcnref}

\end{document}